\newcommand{\CLASSINPUTbottomtextmargin}{25.4mm}
\newcommand{\CLASSINPUTtoptextmargin}{19.1mm}
\newcommand{\CLASSINPUToutersidemargin}{15.8mm}
\newcommand{\CLASSINPUTinnersidemargin}{17mm}
\documentclass[letterpaper,conference]{IEEEtran}
\ifCLASSINFOpdf
\else
\fi
\usepackage{amsmath}
\usepackage{siunitx}
\usepackage{graphicx}
\usepackage{calc}
\usepackage{algorithm}
\usepackage{algorithmic}
\usepackage{color}
\usepackage{array,multirow,makecell}
\usepackage{lineno}
\usepackage{hyperref}      
\usepackage{color}
\usepackage{cite}
\usepackage[font=small]{caption}
\usepackage[font=small]{subcaption}
\usepackage{calc}
\makeatletter
\newcolumntype{P}[1]{>{\centering\arraybackslash}p{#1}}
\renewcommand\figurename{Figure}
\renewcommand{\appendixname}{Annex}
\setcellgapes{1pt}
\makegapedcells

\makeatother
\newcommand{\RN}[1]{%
 \textup{\uppercase\expandafter{\romannumeral#1}}
}
\modulolinenumbers[5]
\begin{document}
\title{{\fontsize{24pt}{24pt}\selectfont
Network Issues in Virtual Machine Migration}\\}
\author{\IEEEauthorblockN{Hatem Ibn-Khedher\IEEEauthorrefmark{1},
Emad Abd-Elrahman\IEEEauthorrefmark{1}, 
Hossam Afifi\IEEEauthorrefmark{1} and
Jacky Forestier\IEEEauthorrefmark{2}}
\IEEEauthorblockA{\IEEEauthorrefmark{1}Institut Mines-Telecom (IMT), Telecom SudParis, Saclay, France. 
\\ Email: \{hatem.ibn\_khedher, emad.abd\_elrahman, hossam.afifi\}@telecom-sudparis.eu}
\IEEEauthorblockA{\IEEEauthorrefmark{2}Orange Labs, Issy Les Moulineaux, France.\\
Email: Jacky.forestier@orange.com}
\thanks{This work is supported by the French FUI-18 DVD2C project: https://dvd2c.cms.orange-labs.fr/public-dvd2c/bienvenue-sur-le-site-du-projet-dvd2c.}}
\maketitle
\begin{abstract}
Software Defined Networking (SDN) is based basically on three features: centralization of the control plane, programmability of network functions and traffic engineering. The network function migration poses interesting problems that we try to expose and solve in this paper. Content Distribution Network virtualization is presented as use case.
\end{abstract}
\IEEEpeerreviewmaketitle

\section{Introduction}
\IEEEPARstart{T}{he} virtualization of resources has addressed the network architecture as a potential target. The basic tasks required in the virtualization substrate are instantiation of new network functions, migration and switching. These basic tasks are strongly dependent on the underlying network configuration and topology in a way that makes them tributary of the network conditions. It means that sometimes it will not be
possible or not recommended to accomplish some virtualization tasks if the network is not presenting the minimum requirements. This brings back many ‘déjà vu’ questions to networks theory but the answers to these questions require an understanding of the new context.

Most of the virtualization architectures such as NFV
included in the SDN paradigm are relying on these tools to
implement their technical solutions.

We consider here the migration of network functions and
we study problems that arise from this dynamic process. We
present a use case for content distribution of multimedia
services.

The rest of this paper is organized as follows; Section II
highlights the common architectures for network resource
virtualization. Section III describes the influence of the main network parameters in the virtualization process. We study network basics such as addressing, quality of service and mobility issues. Simple improvements and technical solutions are presented and demonstrated in Section IV. Section V introduced our virtual CDN (vCDN) use case. Finally, this work is concluded in Section VI.

\section{NFV, SDN and OpenStack}

We investigate the state of the art of the networking
technologies that can meet the virtualization domain. The
main architectures used for that purpose are Network
Functions Virtualization (NFV), Software Defined
Networking (SDN) and OpenStack.

\subsection{Network Functions Virtualization}

Network Functions Virtualization (NFV) \cite{[1]} virtualizes the network equipment (Router, DPI, Firewall...). We will not discuss about hardware. We will rather consider software based NFV architecture. It is a concept that decouples network functions from its underlying hardware. Then, it enables the software to run on virtualized generic environment. Therefore, several virtual appliances can share the single hardware resources.

NFV brings several benefits \cite{[2]} such as reducing CAPEX and OPEX, promoting flexibility and innovation of the virtual network functions already implemented. Moreover, it has been introduced as a new networking facility that poised to amend the core structure of telecommunication infrastructure to be more cost efficient.

In a virtualized environment, the physical resources (CPU,
Storage, RAM…) are emulated and the guest operating system
runs over the emulated hardware (vCPU, vRAM, vStorage …)
as if it is running on its own physical resources. By this
virtualization, all the virtual machines share simultaneously the available resources.

This virtualization is based on deterministic framework
which is constituted of three main components:
\begin{itemize}
\item Host machine: It is the physical machine (physical
server) that owns the hardware’s resources (CPU,
RAM, Storage, Input output interfaces, etc.)
\item Hypervisor: It is the controller of the virtual
machines that controls their instantiation, dynamic
migration, etc.
\item Guest machine: It is the virtual machine or the
virtual appliance that is running and attached to the
hypervisor.
\end{itemize}

The standardization of NFV began with ETSI and some use
cases described in \cite{[3]}. The main use cases found are:
\begin{itemize}
\item Virtualization of the Radio Network Interface:
In this approach we separate the digital function of the radio named Base Band Unit (BBU) from the underlying antenna hardware and distribute the Remote Radio Unit (RRH). The virtualization can be done in a data center that communicates with the
RRH distributed functions through optical back-haul networks (optical fiber) in order to accelerate the resource allocation. Bhaumik et al. \cite{[4]} gave an important result utilizing cloud RAN in which they virtualized the Base Band Units and distribute RRU units. They reported that Cloud RAN can decrease the network load by 22\%.
\item Virtualization of the Home Network: The virtualization of the home network includes the virtualization of the two main components: Residential Gateway and Set-Top Boxes that offer home services (internet access, multimedia service, etc.) to end users. This approach is based on implementing virtualized and programmable software based NFV solution such as: firewalls, DHCP servers, VPN Gateways, DPI Gateways. Then, move them to data centers in order to decrease the cost of devices and increase the QoS.
\item Virtualization of Evolved Packed Core (EPC): In this use case, the virtualization targeted several functions such as: SGW, PGW, MME, HSS, and PCRF \cite{[1]}. The virtual EPC will include all the above functions as software based NFV solutions moved into a cloud EPC. Using this approach we can reduce the network control traffic by 70 percent as described in \cite{[1]}.
\end{itemize}

\subsection{Software Defined Networking}

\begin{figure}[t]
\centering
\includegraphics[scale=0.4]{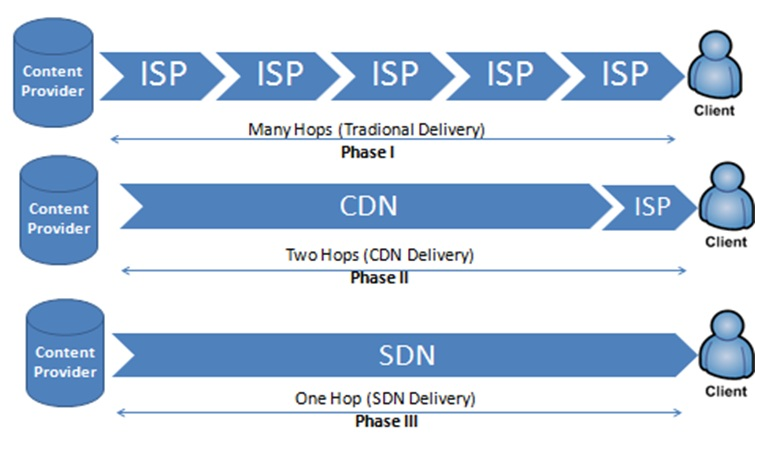}
\caption{Evolution of multimedia application delivery}
\label{1}
\end{figure}

SDN \cite{[5]} is a networking concept that shares the same goals of NFV in the direction of promoting innovation; openness and flexibility. We emphasize that they are independent but complementary as explained in \cite{[6]}.

It is characterized by the separation of control plane and
data plane and consolidates the control plane in one logical
centralized controller to decrease the network overload and
increase enhancement the traffic engineering by adding policy and rules to devices enabled OpenFlow protocol.

It can be considered as a tool for traffic engineering to
enhance the QoS and meet network issues following the
integration of virtualization functions. SDN based on
OpenFlow protocol can be used to direct the forwarding layer
formed by Open Flow enabled devices such as OpenFlow
Switches.

As shown in Fig. \ref{1}, the multimedia service delivery has
evolved three times:
\begin{itemize}
\item Phase I: The operators adopt the Best Effort (BE) in service delivery from content providers to clients according to the available bandwidth and without any regards to delay issues, users’ satisfaction or perception for QoE (i.e. no control for QoS).
\item Phase II: The operators use the CDN solution to cache the
multimedia very near to the clients by using CDN network
acceleration. In this phase, there is a remarkable enhancement
in service delivery while still some restrictions remain
between customers and content providers [15].
\item Phase III: The operators aim to redirect the required services from content providers to selected network locations giving to clients good perception in QoS or QoE measures. In this phase, there is an application layer tunneling defined by SDN (QoS/QoE perception model) [16].
\end{itemize}

Our concern in this paper is to benefit from these
complementarities in order to virtualize and migrate the
multimedia network functions across hosts. Moreover, we aim
to gain the advantages of scalability and cost reduction in a
dynamic controlled service delivery context. We focus in this
work on the mobility issues that arise when achieving the
multimedia network function migration.

\textbf{Mechanism of SDN in Video Service Delivery}

SDN will lead to Software Defined Data Centers (SDDC)
where the roots of streaming points dynamically move. This dynamic adaptation will pass through 4 phases as follows:
\begin{itemize}
\item Resources Virtualization: It concerns the different
resources needed during virtualization including
bandwidth, system (memory, CPU, I/O).
\item Roots \& Links Virtualization: The virtual resources gathered and calculated in real time for the virtual node inserted among different nodes in the original tree of operators’ networks.
\item Network Virtualization: It is a kind of dynamic
networks or nodes on demand to serve as streaming
points.
\item Flows Virtualization: It is a mapping of original root flows to the new elected roots through SDN tunnels between the old root and virtual root to new ones.
\end{itemize}

\subsection{Openstack}

\begin{figure}[t]
\centering
\includegraphics[scale=0.4]{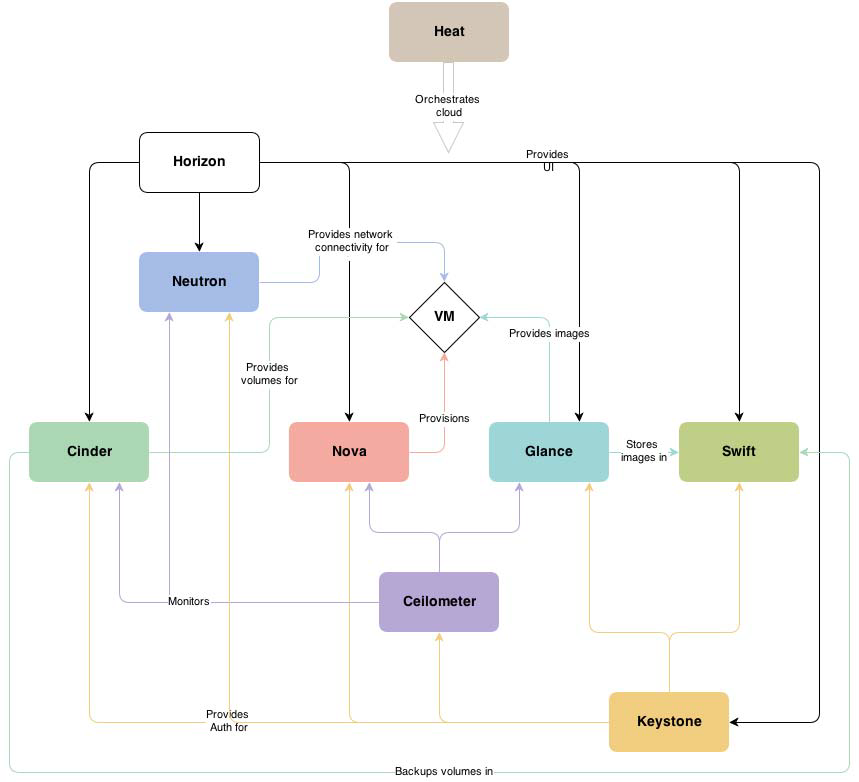}
\caption{OpenStack architecture [7]}
\label{2}
\end{figure}

OpenStack [7] is a collection of services and software. It is an open source project that replaces the Infrastructure as a Service (IaaS) layer in cloud computing domain. It can deal with the two networking technologies that appeared: NFV and SDN. Its conceptual architecture is showed in Fig. \ref{2}.

However the main components of OpenStack are: Nova, Glance and Swift. Glance service is the creator of image disks and Swift project is the manager of the storage in the private cloud (the equivalent of S3 service in Amazon EC2 project).

OpenStack has also an orchestrator module (Heat) that
manages the virtual network functions (VNFs) in NFV based Framework. Their components are well cooperating so as to achieve dynamic instantiation of VMs in the container (VNF) through “Nova” module which is the most important service in this project. It included also Keystone which the Manager of the identity.

\section{Network Constraints in Virtualization} 

We study mobility in the virtualization process. Several basic requirements are necessary before any virtualization procedure can take place. We think that QoS, mobility support and security are among the most important issues to be addressed:

\subsection{QoS} 

This parameter is vital. Any virtualization process
consumes in its transient state a huge amount of resources. Instantiation or migration of a VM requires very high speed links. Slow links do not only render the virtualization very hazardous but can simply make it impossible to accomplish. QoS is required in many situations.
It is to be mentioned that virtual machine migration as of today is not optimized and very trivial approach where everything is copied from the origin to the destination. Hence, we can expect up to 90 per cent of the copied material is un-necessary because it is unchanged. We expect that efforts in live migration will be done to optimize what is being copied. The QoS of the link can hence be a decisive constraint on whether to migrate or not.

\subsection{Mobility} 

Mobility of virtual instances is not a simple task. We can consider moving an NF or a complete instance of a server depending on the desired controller objective. So, from the operating system perspective, there will be a virtual machine, a process, a thread or a session migration as a result of this mobility. As explained before, moving a functionality is required when we want to create a new service in a different location. Off-loading, QoS improvement, interface management [18] etc. may be the reasons for this mobility. When we consider moving a complete virtual machine, it is mainly for load balancing but can also have other reasons such firmware/OS upgrade, instantiation of a new service for a new customer, etc.

\subsection{Security}

Security is an important aspect in VMs migrations. Many attacks could stop the live or offloading migration at any point. So, securing this migration either in single domain or inter multiple domains is mandatory. Tools such as Openstack do not authorize all the possible operations when different domains are involved. This could present a limitation in real deployment.

\section{Networks Issues for Virtualized Network
Function’s Mobility}

Virtualized network functions need for an added network requirement in order to assure session moving, open session and live session migration. For basic knowledge, if a virtual machine that resides on a given network, obviously executed as a process, changes the network; it suffers from the
continuity of the sessions offered to users which have
connected to it. This of course is due to change of VM’s IP. Therefore, we must address this issue to ensure the continuity of the session after the migration of the virtual appliance to a remote host and a remote Hypervisor.

\subsection{Hypervisor Overview}

Hypervisor is the virtual machine monitor that controls virtual machine issue and specially network issues of VM and during the live migration process. It allows to move virtual machines between different domains. Among the famous hypervisors, we cite VMware \cite{[8]}, Xen \cite{[9]}, Qemu [10], and VirtualBox [11] etc.

In our paper, we use two tools. Qemu is first tested as an open source hypervisor
that can monitor the running virtual machine and deal with virtualization tools that can’t be used in another virtualized environment such as: KVM. We developped also a special tool for Optimized migration based on Openstack \cite{[21]} based on linear optimization tools \cite{[19]} and \cite{[20]}

The aforementioned hypervisors enabled VM live migration with continuity of the running service. However, they missed several requirements for assuring live migration and open session. Those requirements are mainly concerning IP mobility when VMs migrate from source host to destination one as it is explained in [12] and [13].

Q. Li at.al [14] introduced that mobility issue can be controlled either by the VM itself, by the host machine (or any intermediate network node) or by the hypervisor. We will focus on those ways of enabling and controlling IP mobility when VMs migrate from one host to another and prove that the best way to control IP mobility is by the hypervisor.

In [17] Kalim et al. introduced another approach to migrate virtual machines between different subnets. In their work, they decouple IP address from TCP transporting layer and enabling transport independent flow in order to solve mobility issues and network configuration problems when migrating VMs. This approach describes the requirements and the mechanisms used. However, it misses the description of live migration scenarios that still remain unclear in their prototype.

In live migration processes any of the VM or the network nodes knows the exact time of migration and the destination host. VM can’t discover that the hypervisor on which it attached to the network has been changed after the migration. Therefore, the mobility problem in live migration can be easily resolved by letting the hypervisor do this work.

\subsection{Basic concepts for MIP enabled live migration}

\begin{figure}[t]
\centering
\includegraphics[scale=0.4]{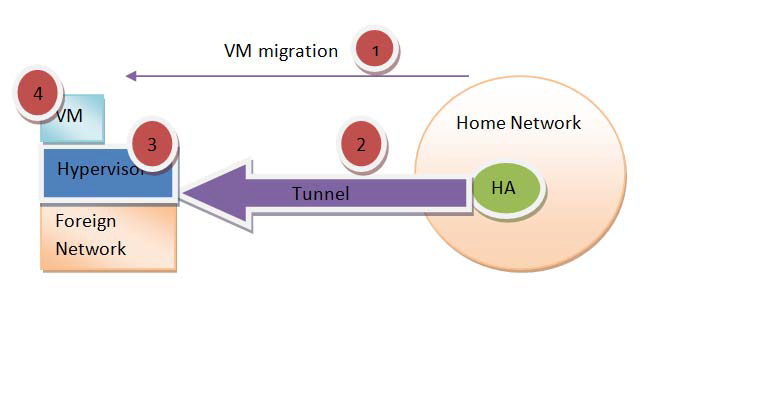}
\caption{Hypervisor monitored Live Migration}
\label{3}
\end{figure}

The basic components in our conceptual architecture are the customer or the client, the home agent and the home node, the foreign agent, the Hypervisor (Qemu in our case) and the correspondent node. We keep the basis component names as in traditional Mobile IP protocol (MIP). Fig. \ref{3} shows the four steps for enabling hypervisor controlled mobile IP for live migration. Firstly, an administrator located at the Home network executes the live migration process. Secondly, a secure Tunnel based on SSH security protocol is created between the home agent and the remote hypervisor on which the migrated VM is
running. Thirdly, the hypervisor software redirects the incoming traffic to the migrated VM. Fourthly, this latter keeps alive its session without any interruption.

\subsection{Networking System Design}

\begin{figure}[t]
\centering
\includegraphics[scale=0.3]{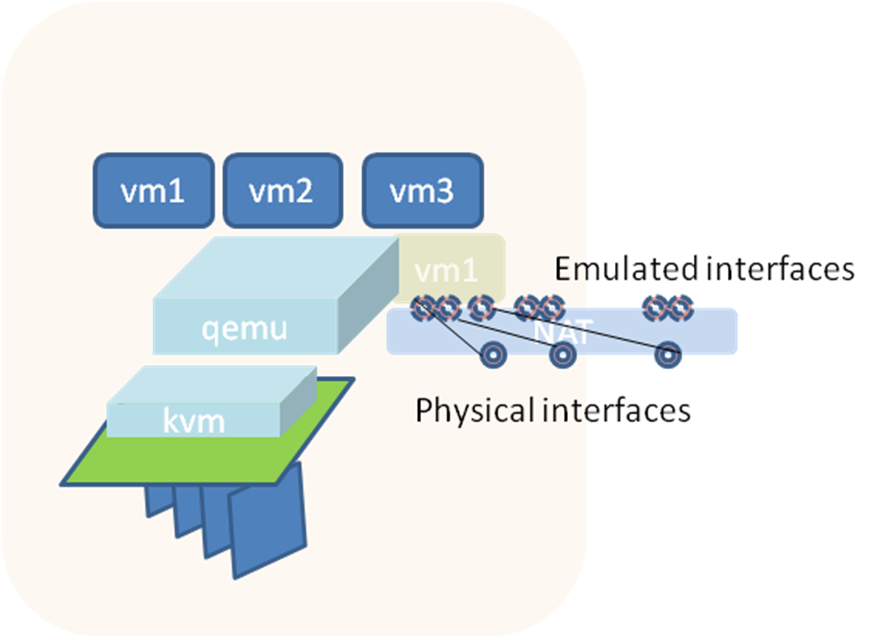}
\caption{VM networking scheme}
\label{4}
\end{figure}

In our basic architecture several virtualization tools were used in order to implement network functions and run them on a virtualized environment. Fig. \ref{4} showed our networking scheme which relies on QEMU/KVM assisted by Libvirt daemon. Those tools manage the running VMs and their issues. The networking scheme used with virtual managers is based on Network Address Translation (NAT) for connecting virtual machines to the physical host. Each VM has emulated interfaces by which they are differentiated

\subsection{Virtual Machine Migration Process}

\begin{figure}[t]
\centering
\includegraphics[scale=0.4]{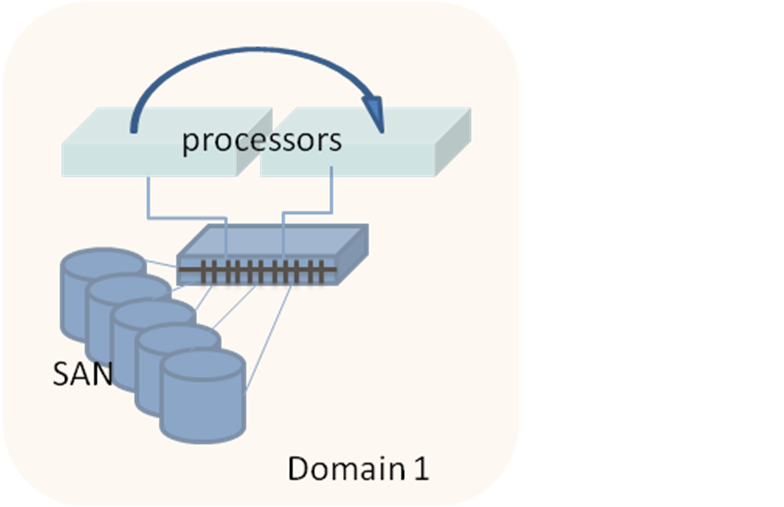}
\caption{Live Migration with shared storage}
\label{5}
\end{figure}

\begin{figure}[t]
\centering
\includegraphics[scale=0.3]{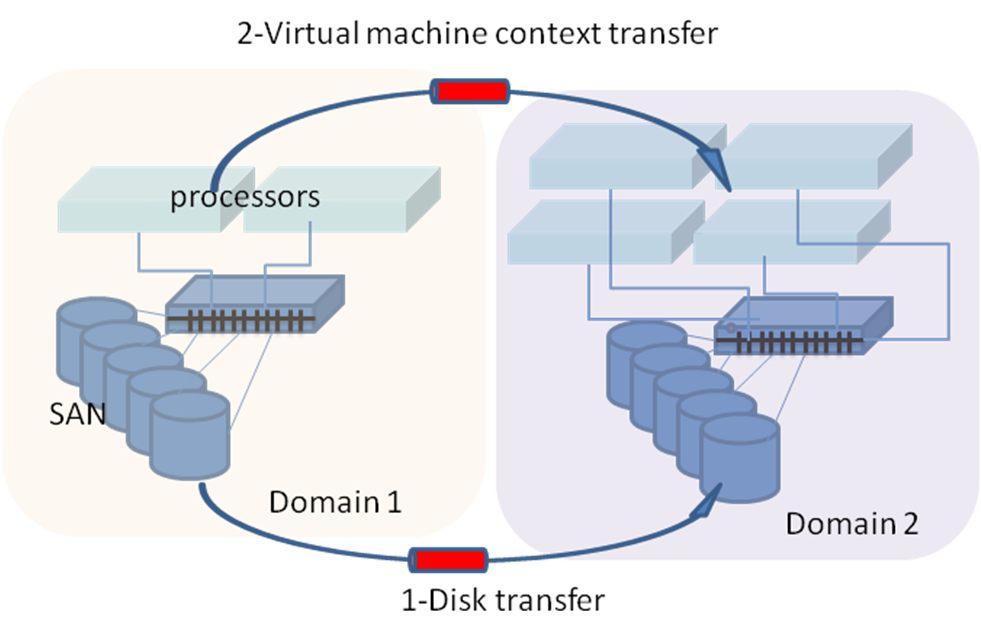}
\caption{Live VM context transfer}
\label{6}
\end{figure}

Virtual machine migration in our scenario enables open session, session moving, content moving and virtual machine moving. We differentiate migration into two types: live VM
migration with shared storage (SAN: storage area network)
and live VM migration with context transfer. Process
migration is still an open problem and session migration does not require virtual solutions.

In Fig. \ref{5}, we show a simple scenario of live VM
migration without context transfer. In this scenario, once the virtual machines are executed, the two physical machines share the same disk image.

In Fig. \ref{6}, we describe a more complex scenario: live migration of a network function with context transfer. Before the migration process, the migrated VM through the hypervisor transfers memory pages and the total disk to the remote host. After the Disk transfer, the hypervisor transfers the virtual machine context (storage, plug-in, packages, libraries, etc.) to the remote host.

\subsection{Evaluation}

\begin{table}[!b]
\caption{LIVE MIGRATION REQUIREMENTS}
\label{1}
{\fontsize{8pt}{8pt}\selectfont
\renewcommand{\arraystretch}{1.3}
\begin{center}
\begin{tabular}{ |p{1cm}| p{1cm}|p{1.2cm}|p{1cm}|p{1cm}|}
\hline  & CPU State & Storage content &Network connection &Memory content \\
\hline With shared Storage& Same context & shared &ARB broadcast &Coping memory pages \\
\hline Without shared Storage&Same context & Transferred, Needed for high link speed &Missed for live session &Coping
memory pages \\
\hline
\end{tabular}
\end{center}
}
\end{table}

Starting from our previous networking scheme based on
Qemu and KVM virtualization tools, we enabled the two way of live migration either with shared disk images (single SAN or domain) or with context transfer (multiple SANs or multi Domain) .

We have focused on network issues that we have faced in live migration in order to more understand the context transfer between the interacted physical and virtual machines. Some system and networking issues are evaluated in this paper. We reported that live migration of running virtual machines needed much more requirements in several migrating components such as CPU state, storage content, network connections and memory content as it showed in Tab. \ref{1}.

We evaluated some experiment measurements basically
related to network parameters such as: link speed that must be higher than 1Gbit/s, live migration time that is too short with shred storage 4s but without session interruption in the case that we moved streaming point from one host to another.

\section{CDN Use Case}

\subsection{CDN}

\begin{figure}[t]
\centering
\includegraphics[scale=0.35]{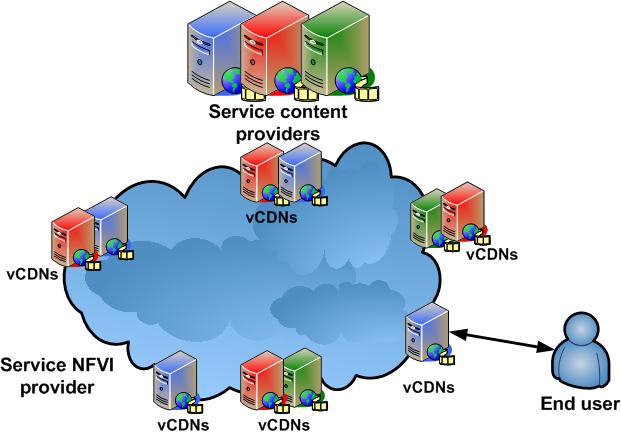}
\caption{vCDN based NFV solution}
\label{7}
\end{figure}

\begin{figure}[t]
\centering
\includegraphics[scale=0.35]{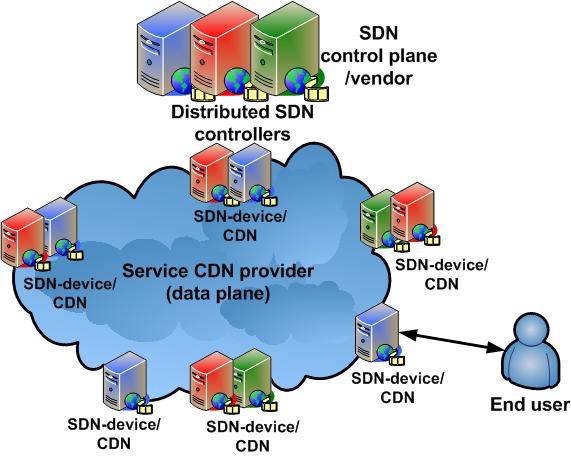}
\caption{vCDN based SDN solution}
\label{8}
\end{figure}

Content Distribution Network (CDN) is a large distributed network deployed worldwide in order to push content on the edge of the network. It was the solution of two major problems that decreased the performance of the network: congestion within the core network and overload at the origin server. It is often used to host static content so that they will be close to the end user when he made his request for online video streaming. Static contents are: images, video, CSS file, scripts files. CDN customers are mainly the Internet Service Providers (ISPs) and Over the Top (OTT) service providers like YouTube and Netflix. All those customers want to push their content to the edge servers over IP technology. A CDN is looking to build a large farm of cache servers deployed worldwide in order to reduce the load on the OTT servers and
provide the following services: Storage (popularity based), Management of cached content (i.e. cache management like replication, replacement, and placement), Distribution of content among cache servers (load balancing), Content Streaming, Fault tolerance (i.e. Redundancy), and Network performance (reduce load).

\subsection{Virtualization of CDN}

CDN federation can be proposed to improve performance. Virtual CDN, vCDN is a virtual solution for enhancing the utilization of the current CDNs.

Most of the internet traffic is delivered through
commercial CDN by allocating some cache servers to
distribute the content to end users. We introduced that this process can be enhanced by three ways of virtualization: classical vCDN based Cloud brokers, vCDN based NFV solution, vCDN based SDN solution and vCDN based NFV and SDN. Among those virtualization ways, we have proposed three new solutions:
\begin{enumerate}
\item vCDN based NFV.
\item vCDN based SDN.
\item vCDN using the merge of NFV \& SDN.
\end{enumerate}

Fig. \ref{7} shows our vCDN based NFV solution. Different services issued from unique or multiple service providers can hence be deployed on virtual machines running as a streaming serves on physical servers (Common-Off-The-Shelf (COTS) Sever). End user’s requests are forwarded to the closest vCDN virtual machine. Several network issues must be resolved to complete NFV and network requirements such as mobility, QoS and QoE.

In Fig. \ref{8}, we describe our second architecture using SDN. In this approach SDN is only used as a traffic engineering or a monitor of edge servers. Through the control protocol OpenFlow (OF) enabled in OpenFlow Switches (OFS), vCDN based SDN can redirect user’s request to the closest surrogate server and enhance then the overall QoS. The
Network Operation System (NOS) hosts the SDN controller to communicate with the forwarding layer via OF protocol. Several control modules and applications can be added to our SDN paradigm to enhance the control plane.

Our third solution: vCDN based NFV and SDN merges
the two former solutions to virtualize and program CDN software. We keep the design of vCDN based NFV solution and we add the SDN paradigm to control the virtual edge servers (vCDN) based on NFV solutions rather than dedicated hardware.

\section{Conclusion} 

In this paper, we highlighted the impacts of main network parameters in the live migration process. We proposed some network virtualization concepts like SDN, NFV and OpenStack. We detailed network issues of VM migration in different scenarios. We concluded that NF context should be transferred for full VNF migration and open session. Virtualized network functions can be dynamically instantiated and migrated in a virtual environment but they need for high requirements to achieve the full context transfer and full virtualization. We explained also a use case of vCDN for multimedia service objective.

\ifCLASSOPTIONcaptionsoff
  \newpage
\fi
 
\IEEEtriggeratref{8}
\IEEEtriggercmd{\enlargethispage{-5in}}

\bibliographystyle{IEEEtranS}
\bibliography{IEEEabrv,references}

\begin{thebibliography}{1}
\bibitem[1]{[1]} B. Han, V. Gopalakrishnan, L. Ji, S. Lee, “Network Functions Virtualization: Challenges and Opportunities for Innovations” in Communications Magazine, IEEE, pp.90-97, February 2015.
\bibitem[2]{[2]} H. Hawilo, A. Shami, M. Mirahmadi, R. Asal, “NFV: State of The Art, Challenges and Implementation in Next Generation Mobile Networks
(vEPC)”, in IEEE Network. Mag., pp.18–26, November 2014.
\bibitem[3]{[3]} The European Telecommunications Standards Institute. Network Functions Virtualization (NFV); Use Cases. GS NFV 001 (V1.1.1), Oct. 2013.
\bibitem[4]{[4]} S. Bhaumik, S. P. Chandrabose, M. K. Jataprolu, G. Kumar, A.Muralidhar, P. Polakos, V. Srinivasan, T. Woo. CloudIQ: A Framework for Processing Base Stations in a Data Center. In Proceeding of MOBICOM 2012, pp.125-136, August 2012.
\bibitem[5]{[5]} A. Basta, W. Kellerer, M. Hoffmann, H. Jochen Morper, K. Hoffmann, “Applying NFV and SDN to LTE Mobile Core Gateway; The Functions Placement Problem”, in SIGCOMM ACM Special Interest Group on Data Communication, pp.33-38, Aughust 2014.
\bibitem[6]{[6]} M. Jammal, T. Singh, A. Shami, R. Asal, Y. Li, “Software Defined Networking: State of The Art and Research Challenges” Computer Networks 72: 74-98, June 2014.
\bibitem[7]{[7]} Web Site: https://www.openstack.org/.
\bibitem[8]{[8]} http://www.vmware.com/fr/
\bibitem[9]{[9]} Web site: http://www.xenproject.org/
\bibitem[10]{[10]} Web site: http://wiki.qemu.org/
\bibitem[11]{[11]} Web site: https://www.virtualbox.org/
\bibitem[12]{[12]} M. Jbella, A. Ben latifa, S. Tabbane, “A Survey Of Live Migration
inVirtual Network Environmrnt (VNE)” in New Technologies for Distributed System (NOTERE), 2010 10th Annual International Conference, pp.351-354, June 2010.
\bibitem[13]{[13]} T. Kondo, R. Aibara, K. Suga, K. Maeda, “A Mobility Management
System for the Global Live Migration of Virtual Machine across
Multiple Sites” in Computer Software and Applications Conference Workshops (COMPSACW), 2014 IEEE 38th International, pp.73-77, July 2014.
\bibitem[14]{[14]} Q. Li, J. Huai, T. Wo, M. Wen, “HyperMIP controlled Mobile IP for Virtual Machine Live Migartion across Networks” in High Assurance Systems Engineering Symposium, 2008. HASE 2008. 11th IEEE , pp.80-88, December 2008.
\bibitem[15]{[15]} E. Abd-Elrahman and H. Afifi, “Moving to the cloud: New visiontowards collaborative delivery for open-iptv”, in ICN 2011, The Tenth International Conference on Networks, 2011, pp. 353–358.
\bibitem[16]{[16]} M.T. Diallo, F. Fieau, E. Abd-Elrahman, H. Afifi, “Utility-based Approach for Video Service Delivery Optimization”, ICSNC 2014: International Conference on Systems and Network Communication, 2014, pp.5-10.
\bibitem[17]{[17]} Kalim, U.; Gardner, M.K.; Brown, E.J.; Wu-chun Feng, "Seamless Migration of Virtual Machines across Networks," Computer Communications and Networks (ICCCN), 2013 22nd International Conference on , vol., no., pp.1,7, July 30 2013-Aug. 2013.
\bibitem[18]{[18]} Sethom, K., \& Afifi, H. (2004, June). Requirements and adaptation solutions for transparent handover between Wifi and Bluetooth. In Communications, 2004 IEEE International Conference on (Vol. 7, pp. 3916-3920). IEEE.
\bibitem[19]{[19]} Hatem Ibn-Khedher, Makhlouf Hadji, Emad Abd-Elrahman , Hossam Afifi and Ahmed E. Kamal. Scalable and Cost Efficient Algorithms for Virtual CDN Migration.
\bibitem[20]{[20]} H. Ibn Kheder, E. Abd Elrahman, H. Afifi, OMAC: Optimal Migration Algorithm for virtual CDN. 23rd ICT. 2016.
\bibitem[21]{[21]} Vios. https://github.com/TelecomSudparis-RST/vIOS

\end{thebibliography}

\end{document}